\documentclass[twocolumn,showpacs,preprintnumbers,superscriptaddress]{revtex4-1}
\usepackage{graphicx}
\usepackage{dcolumn}
\usepackage{bm}
\usepackage{subfigure}
\usepackage{threeparttable}
\usepackage{amsmath}
\usepackage{amssymb}
\usepackage{color}

\begin{document}

\title{Enhancing topology adaptation in information-sharing social networks}

\author{Giulio Cimini}
\email{giulio.cimini@unifr.ch}
\affiliation{Physics Department, University of Fribourg - CH-1700 Fribourg, Switzerland }
\author{Duanbing Chen}
\email{dbchen@uestc.edu.cn}
\affiliation{Web Sciences Center, School of Computer Science, University of Electronic Science and Technology of China, Chengdu 611731, P.R. China}
\author{Mat\'{u}\v{s} Medo}
\author{Linyuan L\"{u}}
\affiliation{Physics Department, University of Fribourg - CH-1700 Fribourg, Switzerland }
\author{Yi-Cheng Zhang} \author{Tao Zhou}
\affiliation{Physics Department, University of Fribourg - CH-1700 Fribourg, Switzerland }
\affiliation{Web Sciences Center, School of Computer Science, University of Electronic Science and Technology of China, Chengdu 611731, P.R. China}
\date{\today}

\begin{abstract}
The advent of Internet and World Wide Web has led to unprecedent growth of the information available. 
People usually face the information overload by following a limited number of sources which best fit their interests. 
It has thus become important to address issues like who gets followed and how to allow people to discover new and better information sources.
In this paper we conduct an empirical analysis on different on-line social networking sites, 
and draw inspiration from its results to present different source selection strategies in an adaptive model for social recommendation.
We show that local search rules which enhance the typical topological features of real social communities give rise to network configurations that are globally optimal. 
These rules create networks which are effective in information diffusion and resemble structures resulting from real social systems. 
\end{abstract}
\pacs{89.75.-k, 89.65.Ef}
\keywords{Information diffusion, Network dynamics, Adaptive model}
\maketitle

\section{Introduction}

The fast development of the Internet has caused the amount of information available to grow dramatically. Therefore, people can hardly find what they are interested in. 
The problem of delivering the right content to the right person has attracted much attention in recent years.  
A possible solution is represented by Recommender Systems \cite{Resnick-CommACM-1997,Herlocker-ACMTIS-2004,Adomavicius-IEEETKDE-2005}, 
which act as personalized information filters by analyzing users' profiles and past activities. 
Techniques used to produce recommendations include Collaborative Filtering \cite{Herlocker-ACMTIS-2004, Linden-IEEEIC-2003}, Bayesian clustering \cite{Breese-ProcCUAI-1998}, 
Probabilistic Latent Semantic Analysis \cite{Hofmann-ACMTIS-2004}, matrix decomposition \cite{Maslov-PRL-2001, Koren-COm-2009}, diffusion and conduction 
\cite{Zhou-PRE-2007, Zhang-PRL-2007, Zhou-PNAS-2010} and many others. However it was recently shown that similarity of users' past activities plays a less important role than social influence:
people value recommendations obtained by abstract mathematical analysis less than those coming from their friends or peers \cite{Sinha-ProcDELOS-NSF-2001}. 
\emph{Social recommendation} has hence emerged as a new approach which makes direct use of the social connections between members of a society \cite{Golbeck-Science-2008}. 
Examples of social recommending implementations include services like Delicious.com, Flickr.com, LiveJournal.com, Youtube.com, FriendFeed.com and Twitter.com,  
where users can select some other users as information sources and \emph{follow} them by importing or receiving respectively their URLs, photos, journals, videos, feeds and microblogs. 
In these systems the information spread from a user to her followers, and eventually to the followers' followers, and so forth. 
This diffusion mechanism resembles the spreading of epidemics or rumors over a network \cite{Moreno-PRE-2004, Zhou-PNS-2006, Lu-NJP-2011}.

A recently proposed news recommendation model \cite{Matus-EPL-2009, Giulio-EPJB-2011, Dong-PhysicaA-2011, Tao-PLOS-2011} mimics the spreading process typical for social systems 
and combines it with an adaptive network of connections. 
In this model, when a user reads a news (or a different kind of content), she can either ``approve'' or ``disapprove'' it. If approved, the news spreads to the user's followers. 
Thus each user receives pieces of news from other users who represent her current \emph{leaders} (i.e. information sources). Simultaneously with the spreading of news 
the leader-follower network evolves with time in order to connect users with similar tastes. A key aspect of this model is hence how to find good sources for each user. 
In \cite{Giulio-EPJB-2011} the authors propose a hybrid strategy for leaders updating based on local search and random off-trap, that is able to efficiently optimize 
the network of connections. The local aspect of the proposed strategy considers the leaders of her current leaders as potential candidates for each user,
increasing in this way the \emph{clustering coefficient} of the network. However this approach leaves aside other potential good candidates. For instance, real life examples 
reveal that a follower of a user is very likely to become a good leader for her too, as suggested by the high value of the \emph{link reciprocity} 
in many information-sharing social networking services.

\begin{table*}
 \caption{Statistics of social networking sites.\label{tab.emph}}
 \begin{ruledtabular}
 \begin{tabular}{l|ccccc}
			& Delicious		& Flickr	 	& LiveJournal		& YouTube		& Friend-Feed		\\
\hline
Date of crawl		& 05-2008		& 01-2007		& 12-2006		& 01-2007		& 09-2009		\\
Number of users		& 854,357		& 1,715,255		& 5,203,764		& 1,138,499		& 513,588		\\
Number of links		& 2,521,187		& 22,613,980		& 76,937,805		& 4,945,382		& 19,810,789		\\
Mean user degree	& 2.95			& 13.18			& 14.78			& 4.34			& 38.58			\\
Reciprocity		& 0.392			& 0.624			& 0.734			& 0.791			& 0.207			\\
Clustering		& 0.161			& 0.165			& 0.255			& 0.077			& 0.146			
 \end{tabular}
 \end{ruledtabular}
\end{table*}

In this paper we first conduct an empirical analysis on different on-line social networks, showing that real social communities are characterized by high values of 
link reciprocity and clustering coefficient. Then, building on the adaptive model introduced in \cite{Matus-EPL-2009}, we pose the following question: 
if the users' choice of the leaders were guided by an automated recommendation method, which methods would lead to good choices? Inspired by our empirical results, we propose 
and compare different local leader updating strategies. Using an agent-based framework, we study the features of the resulting artificial network topology 
from the viewpoint of user' satisfaction, network adaptation and recommendation efficiency. We only rely on local search rules because centralized-search mechanisms are very demanding 
and almost unfeasible for large-scale networks. However we show that this apparent drawback can be overcome by an apt choice of these rules: 
local awareness of the network becomes as effective as global knowledge in producing optimal topologies. 
Moreover we find that an effective local updating strategy actually enhances both reciprocity and clustering coefficient of the network, 
mimicking in this way the users' search of sources (or in general acquaintances) in social networks. 

\section{Empirical analysis}\label{sec_emph}

In this section we extract the features of five different on-line information-sharing social networking sites: 
\emph{delicious.com}, \emph{flickr.com}, \emph{livejournal.com}, \emph{youtube.com} and \emph{friendfeed.com}.
In these systems users form a social network and can share different kind of content---respectively bookmarks, photos, blog articles, videos and feeds. 
Table \ref{tab.emph} gives an overview of the features of the different systems. 
Note that we excluded from the analysis both isolated nodes and self-loops. To describe the networks' topologies, 
we use the formalism of the adjacency matrix (whose entry $a_{ij}$ equals one if there is a link from $i$ to $j$, and zero otherwise) and measure two standard quantities:
\begin{itemize}
 \item \emph{Link reciprocity} ($r$) is the tendency of node pairs to form connections between each other  
and is defined as the ratio of the number of bi-directed links to the total number of links in the network \cite{Wasserman-CUP-1994}:
\begin{equation}\label{eqRec}
r=\frac{\sum_{ij}a_{ij}a_{ji}}{\sum_{ij}a_{ij}}
\end{equation}
 \item \emph{Clustering coefficient} ($c$) measures the tendency of the network to form tightly connected components 
and is defined as the ratio of the number of directed link triangles that exist among a user and her first neighbors 
to the total number of triangles that can exist among these users, averaged over all users \cite{Fagiolo-PRE-2007}:
\begin{equation}\label{eqClu}
c=\frac{\sum_i}{N}\frac{\sum_{jk}(a_{ij}+a_{ji})(a_{ik}+a_{ki})(a_{jk}+a_{kj})}{2[d_i(d_i-1)-2d^\leftrightarrow_i]} 
\end{equation}
where $d_i=\sum_j(a_{ij}+a_{ji})$ and $d^\leftrightarrow_i=\sum_ja_{ij}a_{ji}$ are respectively the total degree and number of bi-directed links of node $i$.
\end{itemize}
These quantities are the simplest and most widely used measures to describe the local link structures of a network and the relationships among \textit{close} nodes 
(in terms of distance on the network). There are other important measures of network's topology, like node's degree or centrality, which do not focus on the interconnections 
between neighboring nodes and were consequently excluded from our analysis.

\emph{Delicious.com}, previously known as \emph{del.icio.us}, is the world-largest online bookmarking website.
Users in delicious.com collect URLs as bookmarks; moreover, they can select other users to be their \emph{leaders} (i.e. information sources) and \emph{follow} them 
by importing their bookmarks. Hence we can naturally represent the delicious.com community by a directed leader-follower network.
To extract the network's structure, we perform a crawl of the user graph by accessing the public web interface provided by the site: 
starting from a user, we follow her outgoing and incoming links to reach other users, and so on. This algorithm is known as breadth-first search (BFS) \cite{Lee-IRE-1961}.
The dataset is being collected since May 2008, and it consists of 854,357 users and 2,521,187 directed links among them; out of these users, more than 99\% belong to the giant component. 

\emph{Flickr.com}, \emph{Livejournal.com} and \emph{Youtube.com} are web services in which users can select other users as friends (leaders, as intended in this paper) 
to get access to their content (photos, blogs and video respectively). The leader-follower networks of these on-line communities were obtained in \cite{Mislove-ProcACM-2007} 
by crawling the large weakly connected component of the corresponding user graphs. The algorithm used for the crawl was again BFS with snowball method \cite{Lee-PRE-2006}: 
the data extraction starts from a set of seed users and then it expands by following the outgoing links of these users to reach new users, and so on. 
\emph{Friendfeed.com} is a microblogging service in which users can share short messages to a list of contacts, who can comment back under the original messages. 
It is also a feed aggregator, importing data from several other services like Twitter, Facebook, YouTube, Flickr and Google Reader. 
The leader-follower network we analyze was crawled in \cite{Celli-ProcLNCS-2010}.

The summary of the results of our analysis is reported in Table \ref{tab.emph}. We immediately notice that both the level of link reciprocity and the degree of clustering 
in all social networks are significantly high---between four and five orders of magnitude larger than the respective values in 
Erd\"os and R\'enyi random graphs \cite{Erdos-PMD-1959} with the same number of nodes and links as the real networks.
This phenomenon has a natural explanation in information-sharing social communities: if two users have common interests each of them can likely provide the other with the right content;
also, people tend to be introduced to other people via mutual friends, increasing the probability that two friends of a single user are also friends. 
In the following sections we will draw inspiration from these observations to define the topology evolution rules of an adaptive model for social recommendation.

\section{Model description}\label{sec_model}

We now briefly summarize the adaptive news recommendation model based on \cite{Matus-EPL-2009,Giulio-EPJB-2011} that will be used for the study of different leader selection strategies. 

The system consists of a network of $U$ users, each connected to $L$ other users (the user's leaders) by directed links. 
The value of $L$ is fixed as users usually follow a limited number of sources. 
Users receive news from their leaders and eventually read and forward them to their followers. In addition, they can introduce new content to the system. 

Evaluation of news $\alpha$ by user $i$ ($e_{i\alpha}$) is either $+1$ (liked), $-1$ (disliked) or $0$ (not read yet).
Similarity of reading tastes of users $i$ and $j$ ($s_{ij}$) is estimated by comparing users' past assessments: 
if $i$ and $j$ evaluated $N_{ij}$ news in common and agreed $A_{ij}$ times, their similarity can be measured in terms of the overall probability of agreement
\begin{equation}\label{eqSim}
s_{ij}=\frac{A_{ij}}{N_{ij}}\left(1-\frac1{\sqrt{N_{ij}}}\right)
\end{equation}
where the term in the parentheses disadvantages user pairs with small overlap $N_{ij}$ (which are more sensitive to statistical fluctuations). 
For $N_{ij}\le 1$, $s_{ij}$ is replaced by a small positive value $s_0$, 
reflecting the fact that even when there are no users' evaluations, there is some base similarity of their interests.
Apart from their ratings, no other information about users is assumed by the model. 

Propagation of news works as follows. When news $\alpha$ is introduced to the system by user $i$ at time $t_{\alpha}$, it is passed from $i$ to her followers $j$ 
with a \emph{recommendation score} proportional to their similarity $s_{ij}$. If this news is later liked by one of users $j$ who received it, it is similarly passed further 
to this user's followers $k$ (with recommendation score proportional to $s_{jk}$), and so on. 
For a generic user $k$ at time $t$, news $\alpha$ is recommended according to its current recommendation score:
\begin{equation}\label{eqRecommenscore}
R_{k\alpha}(t)=\delta_{e_{k\alpha},0}\:(1-\tau^{-1})^{t-t_{\alpha}}\,\sum_{l\in L_k}s_{kl}\:\delta_{e_{l\alpha},1}
\end{equation}
Here $L_k$ is the set of leaders of user $k$ and $\delta$ is the Kronecker symbol: $\delta_{e_{k\alpha},0}=1$ only when user $k$ has not read news $\alpha$ yet 
and $\delta_{e_{l\alpha},1}=1$ only if user $l$ liked news $\alpha$. The sum represents the instance of a user receiving the same news from multiple leaders---recommendation scores 
are summed up in this case, reflecting that a news liked by several leaders is more likely to be liked by this user too. Finally, to allow fresh news to be accessed fast, 
recommendation scores are exponentially damped with time, with $\tau\in(1,\infty)$ being the scale of the damping.

Simultaneously with the propagation of news, connections of the leader-follower network are periodically rewired to drive the system to an optimal state 
where users with high similarity (taste mates) are directly connected. When rewiring occurs for user $i$, the leader with the lowest similarity value ($j$) 
is replaced with a new user ($k$) if $s_{ik}>s_{ij}$. There are different selection strategies for picking new candidate leaders:
\begin{enumerate}
\item\emph{Random rewiring.} $k$ is simply a user picked at random in the network.
\item\emph{Local rewiring.} $k$ is the user in the neighborhood of user $i$ with the maximum value of $s_{ik}$. This mechanism is based on the observation that two users 
who have common acquaintances are likely to have similar interests. As will be discussed in the next section, there are different ways to define such neighborhood.
\item\emph{Hybrid rewiring.} Random rewiring is used in some cases and local rewiring in the others. This mechanism mimics both users searching for friends among friends of friends 
(local rewiring) and having casual encounters which may lead to long-term relationships (random rewiring). 
\item\emph{Global rewiring.} $k$ is the user who maximizes $s_{ik}$ among all users $U$ (this is a local rewiring with the neighborhood being the whole network).
\end{enumerate}

\subsection{Topology adaptation}

The search for new and better information sources is a fundamental feature of many social communities. In the model described above, 
the leader updating procedure is intended to drive the network to an optimal state where users with high similarity are directly connected, 
so that the system is able to efficiently deliver right news to right users. Among the rewiring strategies described above, 
global search mechanism like the \emph{Global rewiring} are very demanding for large-scale networks and also unfeasible without a centralized control, 
whereas, the \emph{Random rewiring} is very inefficient as good new leaders are hardly found by chance. 
One is therefore constrained to use local search rules. The basis of the \emph{Local rewiring} is to define the ``neighborhood'' of a user, 
i.e. a set of close users in the network who stand for possible candidate leaders. 
The choice of a specific neighborhood should be clever enough to allow users to actually find their taste mates. 
For instance, the pool of candidate leaders should not be too large, as in this case the search becomes unmanageable for the system. 
On the other hand, if the neighborhood size is very small (compared to the whole network) the rewiring may stop at a sub-optimal assignment of leaders: the topology evolution 
halts if users' better leaders are at some moment out of the neighborhoods (they can never be reached), meaning that the algorithm got trapped in a sub-optimal state \cite{Giulio-EPJB-2011}. 
A possible solution to this problem is to employ some percentage of randomness in the selection, as in the \emph{Hybrid rewiring}. 
In this way users may happen to get connected regardless of their distance, and the pool of candidate leaders for each user is potentially the whole network. 
In the following analysis we will always make use of a Hybrid rewiring with 10\% of randomness, to exploit mainly the local search but to avoid trapping in a local minimum 
(see \cite{Giulio-EPJB-2011} for a detailed study of the effect of the randomness percentage on the rewiring efficiency).

\subsection{Neighborhood definition}

\begin{figure}
  \includegraphics[width=0.45\textwidth]{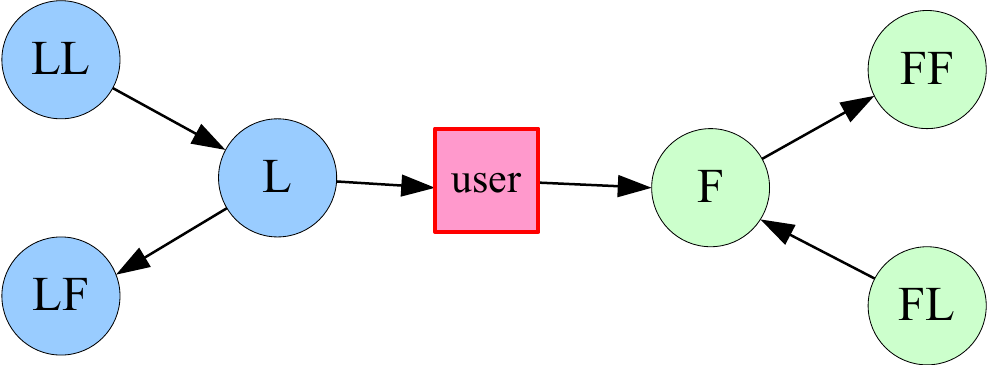}
  \caption{Local network structure of one user. Links' directions reflect how information flows between users.}\label{candidate}
\end{figure}
We shall now define the ``neighborhood'', i.e. the set of candidate leaders exploited by the Hybrid rewiring.
The local network structure from a user's viewpoint is represented in Figure \ref{candidate}. At distance one from the user there are two sets of users: 
her \emph{leaders} ($L$) and \emph{followers} ($F$). $L$ and $F$ form the \emph{first layer} from the user. At distance two, we find four different sets of users: 
her \emph{leaders' leaders} ($LL$), \emph{leaders' followers} ($LF$), \emph{followers' leaders} ($FL$) and \emph{followers' followers} ($FF$). 
These sets form the \emph{second layer} from the user. Notice that the described sets may overlap with each other (e.g. a user can be leading but also following another user).
Given such scheme of the local network structure, we have to consider which of these sets contain potential good leaders for the user. 

Apart from the current set of leaders $L$, the first layer contains a good set of candidates---represented by $F$. 
Indeed if user $i$ is a good leader of user $j$, meaning that $j$ obtains valuable information from $i$, then $i$ and $j$ are likely to have some common interests 
and the similarity between them can be high. Hence also user $j$ can provide user $i$ with the right content and be a good information source for her. 
This assumption is supported by the high value of the \emph{link reciprocity} in many information-sharing social networks (see Table \ref{tab.emph}). 
Including $F$ in the candidate set hence increases the probability of having reciprocal links. However, this set may be too small to be considered alone. 
Therefore we move further to the second layer. The leaders' leaders set ($LL$) was considered in \cite{Giulio-EPJB-2011} where the authors observed that 
since user $j$ obtains valuable information from her leader $i$ and such information may come from $i$'s leaders, then $j$ can have similar interests with $i$'s leaders 
and benefit from following them. Again this assumption is supported by the high value of the \emph{clustering coefficient} in many social networks (Table \ref{tab.emph}).
Analogous considerations lead us to take into account also the $LF$, $FL$ and $FF$ sets. 

In the following sections we will study the behavior of the described model when different rewiring methods are employed. 
When using Hybrid rewiring, we will denote it by the neighborhood that it exploits. 
For instance it will be named as $LL$ if only leaders' leaders are considered as candidates, and $LL+F$ if also followers are included.

\section{Results}\label{sec_result}

For numerical tests of the model, we use an agent-based framework within an artificial network. 
Tastes of user $i$ are represented by a $D$-dimensional binary vector $\mathbf{t}_i$ and attributes of news $\alpha$ by a $D$-dimensional binary vector $\mathbf{a}_\alpha$. 
A taste vector is assigned to each user at the beginning of the simulation, whereas, news have their attributes assigned when they enter the system.
Each vector has a fixed number, $D_A$, of elements equal one (active tastes) and all remaining elements equal zero. 
We always set the system so that all mutually different user taste vectors are present exactly once: $U=\binom{D}{D_A}$ \cite{foot1}. This also means that the taste vectors of two users 
differ at least in two elements. Hence we define as ``taste-mates'' users with exactly two different taste vector elements. 
Opinion of user $i$ about news $\alpha$ is based on the overlap of the user's taste vector with the news's attribute vector
\begin{equation}
\label{eq.O}
\Omega_{i\alpha}=(\mathbf{t}_i,\mathbf{a}_\alpha)
\end{equation}
where $(\cdot,\cdot)$ is a scalar product of two vectors. If $\Omega_{i\alpha}\geq\Delta$ user $i$ likes news $\alpha$ ($e_{i\alpha}=+1$), otherwise she dislikes it ($e_{i\alpha}=-1$). 
Here $\Delta$ is the users' approval threshold. 

Simulation runs in discrete time steps. In each step, an individual user is active with probability $p_A$. 
When active, the user reads and evaluates the top $R$ news from her recommendation list and with probability $p_S$ submits a new news with attributes identical to the user's tastes. 

To build the artificial leader-follower network for the spreading of news, we always start from an initial network configuration with random assignment of leaders to users. 
Then we use the chosen rewiring strategy to update the connections after every $u$ time steps. 

\begin{table}
 \caption{List of parameters used in simulations.\label{tab.para}}
 \begin{ruledtabular}
 \begin{tabular}{lcc}
				& symbol	& value	\\
\hline
Number of users			& $U$		& 3003	\\
Number of leaders per user	& $L$		& 10	\\
Dimension of taste vectors	& $D$		& 14	\\
Number of active tastes per vector& $D_A$	& 6	\\
Users' approval threshold	& $\Delta$	& 3	\\
Probability of being active	& $p_A$		& 0.05	\\
Probability of submitting a news& $p_S$		& 0.02	\\
Number of news read when active	& $R$		& 3	\\
Period of the rewiring		& $u$		& 10	\\
Scale of the time decay		& $\tau$	& 10	\\
Base similarity for users	& $s_0$		& $10^{-3}$
 \end{tabular}
 \end{ruledtabular}
\end{table}

Parameters values used in all following simulations are reported in Table \ref{tab.para}. 

To measure the system's performance we use:
\begin{itemize}
 \item \emph{Approval fraction}, the ratio of news approvals to all assessments: it tells us how often users are satisfied with the news they read 
and is defined as $\sum_{i\alpha}\delta_{e_{i\alpha},1}/\sum_{i\alpha}\delta_{|e_{i\alpha}|,1}$ (here again $\delta$ is the Kronecker symbol). 
In general this quantity can be affected by many factors other than system's performance; however it is a significant measure to consider 
as the main goal of any recommender system is to have users satisfied with what they are recommended---this is a necessary condition for the service to work.
 \item \emph{Average differences}, the average number of vector elements in which users differ from their leaders: it is defined 
as~$\sum_i\sum_{l\in L_i}\|\mathbf{t}_i-\mathbf{t}_l\|_1/UL$ (here $\|.\|_1$ is the 1-norm) and measures how well the network has adapted to users' tastes. 
This is an important quantity to consider as our recommendation method is based on an adaptive network evolution with the aim of having taste-mates directly connected. 
Achieving a low value of average differences is thus a sign of good network adaptation.
\end{itemize}

\begin{figure}
\centering
  \includegraphics[width=0.45\textwidth]{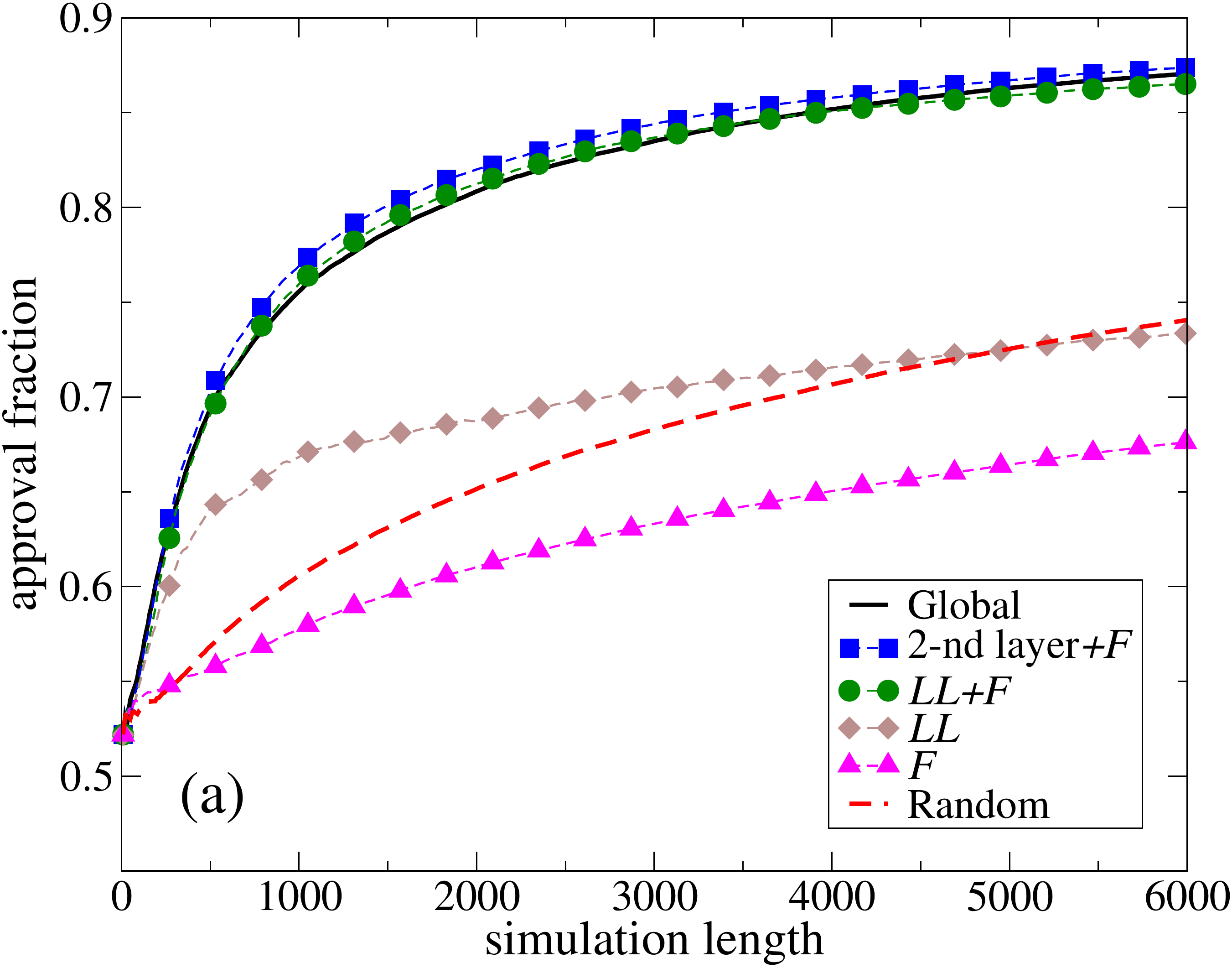}\\
  \includegraphics[width=0.45\textwidth]{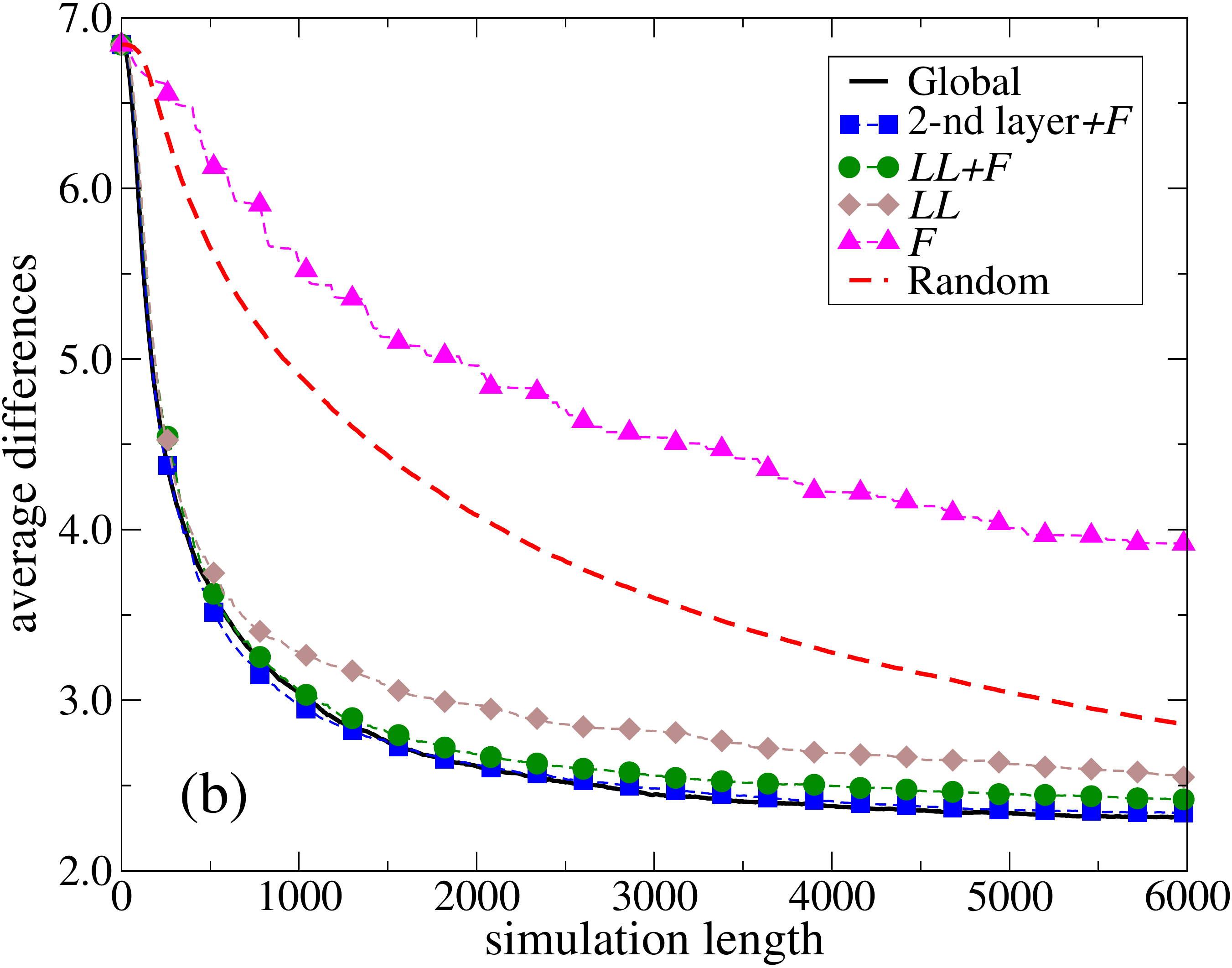}\\
  \caption{Evolution of approval fraction (a) and average differences (b) for different rewiring strategies (single realization of the system). 
Since taste-mates have exactly two different taste vector elements, the lowest possible value of the average differences is two, 
which corresponds to a globally optimal assignment of leaders in the network \cite{foot3}.}\label{fig_AF_AD}
\end{figure}

Figure \ref{fig_AF_AD} shows the approval fraction (a) and the average differences (b) at different times steps of the network's evolution and for different 
definitions of the neighborhood exploited by the Hybrid rewiring. Global and Random methods are shown as benchmarks. 
We observe that any strategy gradually improves both approval fraction and average differences.
As expected, if we limit the pool of candidate leaders to $F$, users are not much satisfied because they can hardly find good information sources. 
This is the result of having considered a very small set (the average number of followers for a user equals $L$). 
If instead we define $LL$ as the neighborhood, as in \cite{Giulio-EPJB-2011}, we significantly improve both users' satisfaction degree and network's adaptation speed. This is because 
the candidate set is much wider in this case---there are on average $L[L-r-(L-1)(c+Lq/2)]$ different leaders' leaders for a users, and this number is much greater than $L$ 
for typical values of $r$, $c$ and $q$ (here $q$ is the probability that four users are linked  in a closed square structure).
To further improve the performance of the system, we expand the candidate pool to $LL+F$. With this definition of the neighborhood we promote at the same time 
the reciprocity and the clustering coefficient of the network, obtaining a surprising effect: both approval fraction and average differences become as good as the ones 
obtained by the Global rewiring, i.e. by considering the whole network as the candidate leader set. In other words, such a small local scale turns out to be as representative 
as a whole-network scale. Hence further expanding the candidate set to the whole second layer (i.e. $LL+LF+FL+FF+F$, 
or 2\textsuperscript{nd}layer$+F$ for shortness) does not bring to any substantial improvements. 
\begin{table*}
 \caption{Statistics of artificial networks at equilibrium for different rewiring strategies. 
The numbers in parenthesis are the errors on the last significant figures. 
a.f. and a.d. stand for approval fraction and average differences, respectively. 
The reference values for $r$ and $c$ in our setting are $r_*=0.208$ and $c_*=0.053$.\label{tab.stat}}
 \begin{ruledtabular}
 \begin{tabular}{c|cccccc}
		&	Global	&	2\textsuperscript{nd}layer$+F$&	$LL+F$	&	$LL$	&	$F$	&	Random	\\
\hline																									
a.f.		&	0.870(4)&	0.870(4)&	0.866(2)&	0.735(6)&	0.670(5)&	0.739(1)\\
a.d.		&	2.35(2)	&	2.35(3)	&	2.43(1) &	2.54(6)	&	3.85(4)	&	2.86(2)	\\
$r$		&	0.63(2)	&	0.65(2)	&	0.61(1)	&	0.13(1)	&	0.85(1)	&	0.12(0)	\\
$c$		&	0.24(2)	&	0.27(2)	&	0.25(1)	&	0.17(1)	&	0.01(0)	&	0.03(0)	\\
$1-\alpha$	&	0.96(2) &	0.96(1)	&	0.97(1)	&	0.96(1)	&	0.81(0)	&	0.82(0)	\\
$1-\beta$	&	0.10(4)	&	0.10(3)	&	0.07(2)	&	0.04(3)	&	0.28(0)	&	0.32(0)	
 \end{tabular}
 \end{ruledtabular}
\end{table*}

We also measure the values of link reciprocity and clustering coefficient in the network. 
We firstly introduce two reference values for $r$ and $c$. In the initial random network the average probability to find a reciprocal link between two connected vertices 
is simply equal to the average probability of finding a link between any two vertices, which is given by $(UL)/[U(U-1)]=L/(U-1)$. Hence we have $r_0=L/(U-1)$. 
This equality also holds for the probability to find a closed link triangle between three users, hence $c_0=L/(U-1)$. 
Instead if the network is in a structure-less optimal configuration where each user has randomly chosen $L$ of her $N=D_A(D-D_A)$ taste-mates as leaders, 
then the value of the reciprocity becomes $r_*=L/N$. Besides in this network state the probability that two taste-mate neighbors of a user are also taste-mates with each other 
is given by $(D-2)/(N-1)$. To show this, consider two taste mate users: there are $(D_A-1)+(D-D_A-1)=D-2$ other users who are taste mates with both of them 
and $(D_A-1)(D-D_A-1)=N-D+1$ who are taste mates with only the first of them. The clustering coefficient is given by the above-mentioned probability 
conditional to the existence of a link: $c_*=[(D-2)/(N-1)][L/N]$.

The stationary values of $r$ and $c$ are reported in table \ref{tab.stat}. 
For any rewiring method, the reciprocity coefficient increases with respect to its initial value $r_0$. 
As expected, the final value of $r$ is very high with $F$ (by construction, $F$ promotes reciprocity) and low for Random and $LL$. In the latter cases, 
$r$ becomes close to $r_*$. The other methods achieve similar values of $r$, which are comparable with the ones of real social networks. 
The clustering coefficient shows an opposite trend: starting from $c_0$, it becomes very large with $LL$ (by construction), whereas, it remains 
quite small for Random and $F$---converging to values close to $c_*$. For the other methods $c$'s stationary values are again comparable 
with the ones of real social networks. 

The fact that the values of $r$ and $c$ in our artificial network are high is not surprising, 
as the local search strategies are built with the aim of enhancing these quantities. 
Since our main purpose is not to model the evolution of a real social network, we only provide a quantitative comparison for reciprocity and clustering values in real and artificial systems: 
In the adaptive model each user has $L=10$ connections, which is of the same order of magnitude of the average degree of a user in the studied real social communities (Table \ref{tab.emph}), 
hence the link densities in these systems are comparable. 
The values of $r$ and $c$ in our artificial system (which are not set by hand but result from the topology adaptation) and in real networks also are of the same magnitude. 
We can conclude that these networks, despite having different size, exhibit similar local link patterns. 

At last, we discuss the efficiency of the modeled recommender system. When making recommendations, it is possible to fall into two different kinds of error: 
recommending content that users wouldn't like, and not recommending content that users would like. These errors are known respectively as of type I (\emph{false positives}) 
and of type II (\emph{false negatives}) \cite{Neyman-Biometrika-1928}. To complete the picture, \emph{true positives} are recommendations of content that users would like, 
and \emph{true negatives} are lacks of recommendation of content that user wouldn't like. Note that false positives upset users but false negatives do not 
(i.e. a type I error has more serious consequences than the other), hence a good recommendation engine should mainly reduce false positives. 
We further introduce the \emph{specificity} ($1-\alpha$) and the \emph{sensitivity} ($1-\beta$) of the recommendation system as the ability to avoid respectively 
false positives and false negatives \cite{foot2}: $$1-\alpha=\frac{\mbox{TN}}{\mbox{TN+FP}}\;\;\;\;\; , \;\;\;\;\;1-\beta=\frac{\mbox{TP}}{\mbox{TP+FN}}$$
where TP, TN, FP and FN are respectively the number of true positives, true negatives, false positives and false negatives. 
To measure these quantities in our artificial setting, 
we define $\alpha$ as the average number of wrong recommendations for a news over the number of users who might dislike this news---given by 
$\sum_{k=0}^{\Delta-1}\binom{D_A}{k}\binom{D-D_A}{D_A-k}$, and $1-\beta$ as the average number of good recommendations for a news 
over the number of users who might like this news---given by $\sum_{k=\Delta}^{D_A}\binom{D_A}{k}\binom{D-D_A}{D_A-k}$. 

Table \ref{tab.stat} also reports the stationary values of specificity and sensitivity for the recommender system when different source selection strategies are employed.
Specificity is remarkably high for all methods, especially for the best performing ones, hence the number of false positives in the system is very low. 
Sensitivity shows instead an opposite trend: Random and $F$ updating strategies are the best performing now. 
We see that the effort of reducing one type of error results in increasing the other type, as it generally happens in statistical tests. 
In our case the reason behind this phenomenon is the presence of tightly connected components in the system: in a highly clustered network 
news have few paths to spread far from the users who post them (and the spreading process takes long time), hence they tend to remain localized. 
As a consequence, few users receive a news but almost all of them like it. 
When clustering is low, a news has more spreading directions, hence it can reach many users but more of them eventually dislike it. 
However we are mainly interested in having a recommender system with high specificity, and in this respect simple local strategies like $LL+F$ again perform 
at the same level of global search in generating optimal network structures for recommending and sharing information. 

\section{Conclusion}

How to recommend the right content to the right person and which are this person's favorite information sources are fundamental questions in the age of information overload. 
In this work we exploited a recently proposed news recommendation model which combines similarity of users' past activities and social relationships to obtain recommendations, 
and which mimics the spreading process typical for social systems where the network of connections continually evolves with time \cite{Matus-EPL-2009}. 
The topology evolution is intended to provide users with newer and better information sources. 
Since global optimization of the users' connections is computationally prohibitive for a large system, a key issue of the model is where to find good new leaders for users. 
Taking real life as inspiration, we designed different local leader search strategies which mimic the users' search of sources in real social communities. 
We then studied the resulting evolution and properties of our artificial system and showed that with these local search rules 
the users' community can self-organize into optimal topologies which are equivalent to the ones that can be generated by global knowledge of the system. 
Indeed the resulting artificial networks have high values of reciprocity and clustering, as similar to the real information-sharing social communities 
studied in section \ref{sec_emph}. Therefore our automated abstract rules help to create networks which not only are effective for the spreading of information 
but also resemble structures resulting from real human activity.

We would like to remark that our main goal is not to model the evolution of a real network. 
The recipes we proposed for the optimization of local network connections are instead a valuable tool which may find application in many systems 
other than the recommendation model presented in this paper---among which social and p2p networks are just a few examples. 
The observed features of the studied social networks suggest that these local rules for topology adaptation also stand for possible mechanisms 
underlying the microscopic evolution of real social communities \cite{Leskovec-ProcACM-2008}.

Finally we recall that the adaptive recommendation model presented in this work is new as it combines similarity-based and social recommendation with the spreading of news. 
An agent-based framework as the one we used in this work represents an ideal playground for testing and of the model and can significantly contribute 
to our understanding of the system \cite{Miller-PUP-2007}. In past works it was used to assess the filtering efficiency of the system \cite{Giulio-EPJB-2011} 
and to study the formation of scale-free leadership structures \cite{Tao-PLOS-2011}. In this work it was employed to test and validate the proposed strategies for network adaptation, 
a task which would have been hard to achieve in a field study with real users. 
Agent-based modeling is also useful for comparing our method with other recommendation techniques (Appendix \ref{appA} and \cite{Matus-EPL-2009}).
In addition to our efforts to have robust simulation results (Appendix \ref{appB}), it still would be beneficial to have direct empirical input for user behavior.
We are working on an on-line implementation of the system and we expect to have in the close future relevant data about real user actions, 
which will be valuable material for future research.
 
The basic feature of our model is recommendations coming from sources selected by the users themselves. 
As there is increasing evidence that users are more inclined to buy products \cite{Leskovec-ACMToW-2007} or like contents \cite{Zhou-ACMWSDM-2012} recommended by friends or trusted peers 
than by others or by abstract mathematical algorithms, our new adaptive recommendation method is a promising candidate to be employed in various social and commercial applications.

\begin{acknowledgments}
This work was partially supported by the Future and Emerging Technologies programmes of the European Commission 
FP7-ICT-2007 (project LiquidPublication, grant no. 213360) and FP7-COSI-ICT (project QLectives, grant no. 231200), 
by the Swiss National Science Foundation (grant no. 200020-121848), by the National Natural Science Foundation of China (grant nos. 60973069, 90924011 and 60903073) 
and by the International Scientific Cooperation and Communication Project of Sichuan Province in China (grant no. 2010HH0002). 
G. Cimini and D. Chen contributed equally to this work. 
\end{acknowledgments}

\appendix

\section{Comparison with other \\recommender systems}\label{appA}

In the adaptive model proposed here, the recommendation process is based on the computation of users' similarity scores from their past assessments. 
In this respect, the model is similar to a widely-adopted recommendation technique: memory-based collaborative filtering (CF) \cite{Herlocker-ACMTIS-2004, Linden-IEEEIC-2003}. 
The idea behind CF is to make predictions about the interests of a user by collecting past preferences from the community the user belongs to, 
with the underlying assumption that those users who agreed in the past tend to agree again in the future. In particular, in memory-based CF 
the recommendation score for an object is computed as a weighted average of ratings given by other users with weights proportional to the similarity between users~\cite{Su-AAI-2009}:
\begin{equation}\label{eqCF}
R_{i\alpha}^{\mbox{\tiny{CF}}}=\frac{\sum_{j=1}^Us_{ij}\,e_{j\alpha}}{\sum_{j=1}^Us_{ij}}
\end{equation}
where $e_{j\alpha}$ is user $j$'s opinion about object $\alpha$.
An alternative approach is to consider for each user only the top-L more similar users and solely use their assessments to compute the recommendation scores~\cite{Su-AAI-2009}: 
\begin{equation}\label{eqtopLCF}
R_{i\alpha}^{\mbox{\tiny{top-L CF}}}=\frac{\sum_{j\in \mbox{\tiny{top-L}}_i}s_{ij}\,e_{j\alpha}}{\sum_{j\in \mbox{\tiny{top-L}}_i}s_{ij}}
\end{equation}
Recently, diffusion-based methods were employed to efficiently extract the similarity between objects or users~\cite{Zhou-PRE-2007, Zhang-PRL-2007}. 
Among these methods, which can be considered as extensions of CF, the \textit{probabilistic spreading} (ProbS) algorithm was shown to outperform 
other standard memory-based CF techniques~\cite{Zhou-PRE-2007, Zhou-PNAS-2010}.
ProbS is based on the bipartite network of $U$ users and $O$ objects, where a link between user $i$ and object $\alpha$ exists if $\alpha$ was collected by $i$
(meaning that the relative entry of the bipartite adjacency matrix $b_{i\alpha}$ equals one, and zero otherwise). 
For each user $i$, ProbS assigns objects an initial level of resource $f^i_\beta=b_{i\beta}$ and then redistributes it to obtain recommendations for uncollected objects via:
\begin{equation}\label{eqPS}
R_{i\alpha}^{\mbox{\tiny{ProbS}}}=\sum_{\beta=1}^O\,\frac{1}{k_\beta}\sum_{j=1}^U\frac{b_{j\alpha}b_{j\beta}}{k_j}f^i_\beta
\end{equation}
We will use these three approaches for testing of our adaptive model---see \cite{Matus-EPL-2009} for a comparison with simpler recommendation methods.

While our adaptive model and these recommender systems are intrinsically different, with the time dimension embedded in the first but totally missing in the others, 
we can still compare the performance of the various methods in the agent-based framework as follows. We let the adaptive system run until it sufficiently approaches equilibrium, 
then when we freeze the evolution and store the current set of users' recommendation lists produced by the adaptive model. 
Since in the artificial setting we have the luxury of knowing the opinion of each user about each news, we can check how many of the news in a user's list 
would be liked by the user (we take into account only the top $S$ places, as real users usually consider only the top recommendations). 
Averaging these values over all users we obtain the mean precision $p$ of the recommendation process \cite{Herlocker-ACMTIS-2004}. A still better perspective is given by considering 
$p$ relative to the precision of random recommendations $p_r$ (which can be easily evaluated in our agent-based setting). This gives the precision enhancement $p/p_r$.
To assess the performance of the other methods, we use all users' assessments resulting from the system's evolution 
to compute the similarities between users \cite{foot4} and the consequent recommendation scores by CF and top-L CF, 
and to build the user-news bipartite network \cite{foot5} for computing the recommendation scores by ProbS. 
From these scores we obtain for each user a recommendation lists, and then we proceed as described before to obtain the precision enhancement values.

However there is a fundamental difference between the recommendations by the adaptive model and the ones by the other algorithms. 
While in the first the time evolution and the damping mechanism allow only recommendations of fresh news, in the others there is no such restriction. 
Hence CF and ProbS often recommend old news which are in principle liked by users but which are not of topical interest. This represents a major drawback, 
especially for a news recommendation engine. To overcome this flaw and to have a fair comparison, we can modify recommendation scores 
by applying the same damping mechanism as in the adaptive model (\ref{eqRecommenscore}):
\begin{equation}
\label{eqRSdump}R_{i\alpha}\leftarrow R_{i\alpha}\;(1-\tau^{-1})^{t-t_\alpha}
\end{equation}

\begin{figure}
\centering
  \includegraphics[width=0.475\textwidth]{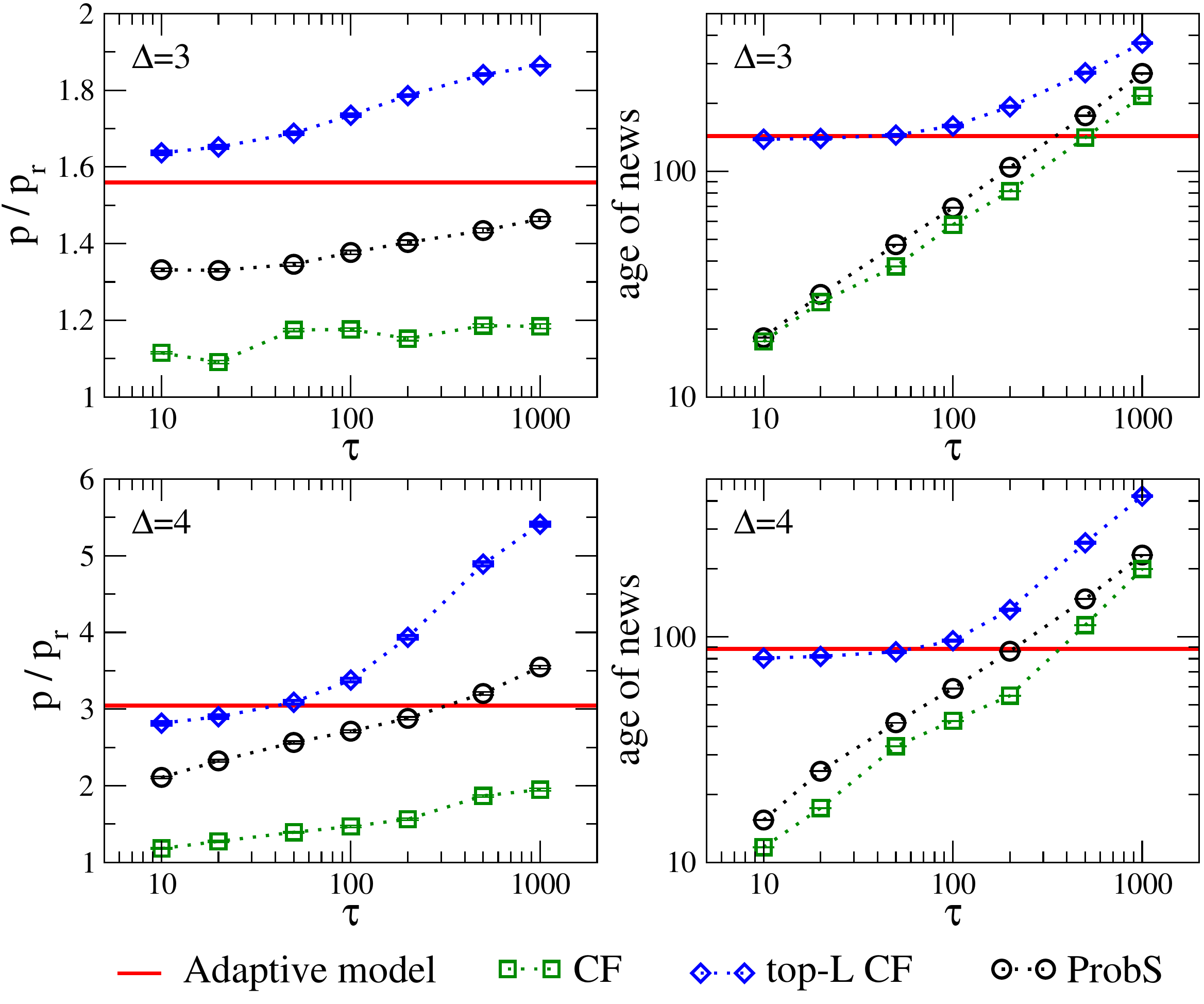}\\
  \caption{Precision enhancement of the recommender system (left) and average age of the recommended news (right) by the adaptive model (optimized by the $LL+F$ rewiring with $\tau=10$) 
and by CF, top-L CF ($L=10$) and ProbS algorithm.  We have used $S=20$. 
Upper and lower plots refer to different system settings in which users are less and more demanding ($\Delta$ equal to 3 and 4, respectively).\label{fig_prec}}
\end{figure}

Figure \ref{fig_prec} shows the precision values and the average age of recommended news ($t-t_\alpha$) by the adaptive model and the other recommender systems. 
We observe that the adaptive model substantially outperforms CF and ProbS unless one is allowed to recommend also very old news ($\tau\gg10$). 
This is because these methods make use of all available information and are able to compute the recommendation score for almost every user-news pair. 
Then if the time decay is weak these methods tend to recommend old and already popular content, 
as typical of standard recommendation techniques which are usually ineffective when dealing with recently introduced object, about which there are few users' feedbacks. 
On the other hand if the decay is strong the freshness of the news becomes dominant over their recommendation scores, to the detriment of precision. 
Our model can instead produce very precise recommendations for fresh news. This is because the memory-based component (i.e. the similarity estimation) 
serves to select a few best information sources for each user, who act as her information filters. Then, given the resulting social network, users only get very precise recommendations, 
among which the most fresh emerge thanks to the damping mechanism. The importance of considering only recommendations coming from taste-mates is also reflected by the performance of top-L CF 
(which is very close to the one of the adaptive model for similar values of $\tau$---the top-L approach is in fact a global rewiring at each step followed by the filling of users' 
recommendation lists). The advantage of our model with respect to this last approach is that the local search rules for leaders' selection 
and the recommendations automatically resulting from the news spreading process overcome the limitations of top-L CF related to scalability and real-time performance \cite{Karypis-IKM-01}.
In addition our model is naturally open to manual selection of leaders by users themselves, 
exploiting in this way social recommendation with all its benefits \cite{Sinha-ProcDELOS-NSF-2001,Golbeck-Science-2008}.

\section{Assumptions \\of agent-based modeling}\label{appB}

The agent-based model introduced in the previous section was very helpful for understanding how our adaptive system behaves. 
At the same time the complexity of its assumptions is such that it is not clear if the reported behavior is general or if it is parameter-dependent. 
Here we discuss the robustness of our results with respect to individual assumptions.

\begin{figure}
\centering
  \includegraphics[width=0.425\textwidth]{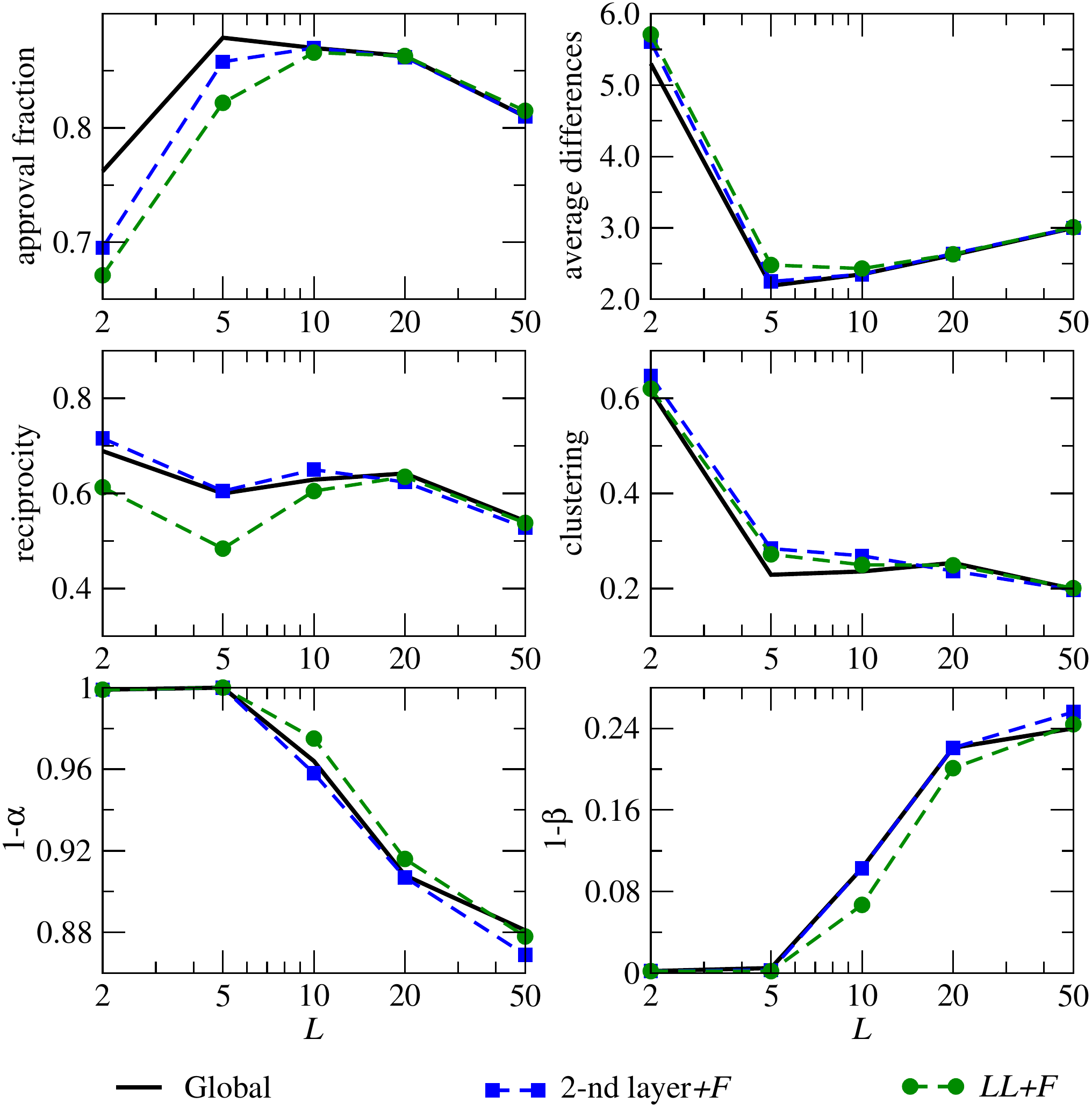}\\
  \caption{Stationary values of approval fraction, average differences, reciprocity, clustering, specificity and sensitivity of the system for various numbers of leaders per user ($L$).}\label{fig_app}
\end{figure}

The first important point to clarify is that the observed features of the system do not depend on the number of users $U$. 
We run simulations of a 15-times bigger network with $U=\binom{18}{8}=43758$ and $\Delta=4$ and observe that the simple $LL+F$ has again the same performance as the global search.
Another important parameter is the number $L$ of leaders per user, which regulates how many news a user receives and in general how many spreading directions there are in the system. 
Figure \ref{fig_app} shows how the system behaves for different values of $L$. 
When $L$ is small news propagate slowly and hardly reach their intended audience. This is why the system is not able to adapt well 
(users' rating histories do not overlap enough), and despite the fact that the rate of false positives is very low users end up reading also news that do not fit their interests. 
On the other hand if $L$ is large news get a large audience but the system loses its filtering efficiency: there are many false positives that cause users to be unsatisfied. 
For a given system size, one can hence set $L$ to a value which gives the best compromise between $1-\alpha$ and $1-\beta$, which also results in a maximum of the approval fraction 
($L\simeq 10$ for local search methods, in our case).

Moving further, $R$, $p_A$ and $p_S$ regulate the amount of news which spread over the network at a given time and the rate of news consumption by users. 
We set them in order to always have enough content in the system, so that users have access every time to diverse news and actually read news forwarded by others. 
However we observe that the system works reasonably well unless one of two possible situations arises. 
If the news circulating in the system are too few, users who are hungry for information will end up reading content that they won't like: 
no recommender system can work if there is too little choice. On the other hand, if the available news are too many then each user will read different news: 
there is too little information and thus the similarity values cannot be always evaluated properly---if such a situation occurs in reality, 
one might employ additional filtering to prevent these negligible effects. However also in this case neighboring users are likely to have some overlap of reading histories, 
hence local search methods can still be successfully employed.

$\tau$ is the scale for the decaying of recommendation scores with time, 
and has to be set such that news can survive for enough steps to spread widely, allowing similarity values to be assessed reliably.
Finally, $\Delta$, $D$ and $D_A$ determine how much users are demanding and how wide are their scopes of interest (see \cite{Matus-EPL-2009} for a discussion on how the system's performance 
relates to these quantities), whereas, $u$ is the frequency for link rewiring, which is set to an optimal value for having both effective recommendations 
and low computational complexity \cite{Dong-PhysicaA-2011}.


\begin{thebibliography}{99}
\bibitem{foot3} As for each user there are $N=D_A(D-D_A)$ possible optimal leaders who differ exactly in two taste vector elements, 
the optimal state is unique only if $L=N$. If $L<N$, there are different possible optimal states, which are equivalent in term of quality of leaders' assignment 
but may differ under other properties like $r$ and $c$. Initial conditions and users' dynamics determine the particular equilibrium state. 
If $L>N$, the average differences are greater than two even in the optimal state.
\bibitem{foot2} Other measures of system's performance (like precision and accuracy) exists and can be derived from $1-\alpha$ and $1-\beta$ in our artificial setting.
We chose to focus on specificity and sensitivity as they show the strongest dependence on how the neighborhood is defined and hence on the resulting local network topology.
\bibitem{foot4} For the similarity estimation, we adopt the same definition as in the adaptive model---eq. (\ref{eqSim}).
\bibitem{foot5} We only consider binary rating for building the bipartite network: $b_{i\alpha}=1$ if $e_{i\alpha}=1$, and $b_{i\alpha}=0$ both if $e_{i\alpha}=0$ and $e_{i\alpha}=-1$.
\end{thebibliography}
\end{document}